# Statistical methods: Basic concepts, interpretations, and cautions

For: *Handbook of Epidemiology*, 3rd edn. I. Pigeot and W. Ahrens, eds.

Sander Greenland, Professor Emeritus

Department of Epidemiology and Department of Statistics

University of California, Los Angeles

Please report errors and ambiguities to lesdomes@ucla.edu

25 August 2025

**Abstract**. The study of associations and their causal explanations is a central research activity whose methodology varies tremendously across fields. Even within specialized subfields, comparisons across textbooks and journals reveals that the basics are subject to considerable variation and controversy. This variation is often obscured by the singular viewpoints presented within textbooks and journal guidelines, which may be deceptively written as if the norms they adopt are unchallenged. Furthermore, human limitations and the vastness within fields imply that no one can have expertise across all subfields and that interpretations will be severely constrained by the limitations of studies of human populations. The present chapter outlines an approach to statistical methods that attempts to recognize these problems from the start, rather than assume they are absent as in the claims of "statistical significance" and "confidence" ordinarily attached to statistical tests and interval estimates. It does so by grounding models and statistics in data description, and treating inferences from them as speculations based on assumptions that cannot be fully validated or checked using the analysis data.

## Introduction

Conventional statistical methods assume that, within measured covariate levels, the data come from surveys with purely random selection from a sharply defined population, or from experiments with purely random treatment assignment in which the researcher has complete control and understanding of every important aspect of those who entered the analysis and how they were treated. The methods thus provide logically sound inferences only in ideal cases, rather than correct answers for our actual imperfect applications. They are however widely applied and misinterpreted in studies that fall far short of their assumptions, as when actual participation and treatment are not random. These problems are only partially accounted for by discussions of study limitations; occasionally they are further analyzed with speculative models for the non-random (systematic) aspects of the causal processes generating the data.

The present chapter will assume the reader has had an introduction to basic statistical methods, and will show how to reinterpret those methods in ways that directly acknowledge these problems based on ordinary-language description of the context and statistics. This is opposed to the usually misinterpreted jargon of “statistical significance” and “confidence”, which assume implicitly that there are no uncontrolled problems with the study. It begins with description of the data. It then proceeds to delineate the contextual factors that may help to explain why study data appeared as they did. These include factors or covariates that can affect selection, treatment, measurement, item missingness, or loss from observation (censoring). Without such covariate information, we have no basis for interpreting statistical outputs as anything more than data summaries.

**Contextual narratives versus statistical formulations**

To facilitate correct use of statistical methods, the study questions the methods are supposed to answer need to be formulated in a clear contextual narrative phrased in ordinary language. This narrative should make the questions apparent even to non-technical observers (e.g., "how much will this treatment affect blood pressure?"). These questions must then be translated into more precise quantities and more specific persons in the study (e.g., "how much did this treatment alter the mean diastolic and systolic blood pressures among the patients in the study?"). Also needed are detailed descriptions of the protocol for participant selection and retention in the study, and of those selected for the study (including what is known about those who refused and those lost after recruitment).

Due to practical limitations, the mathematics of statistical methods will typically involve artificiality and oversimplification, omitting important details of the study's conduct, analysis, and its targets, and thus producing overconfident results (Gelman and Loken 2014a, b; Gelman 2016; Greenland 2017; Amrhein et al. 2019a, 2019b; Amrhein and Greenland 2022; Rafi and Greenland 2020; Greenland et al. 2022). Furthermore, most methods focus on average outcomes, but doing so can obscure a situation in which treatment helps some patients and harms others; it also misrepresents effects in those who did not participate in the study. The statistical results may then invite inappropriate generalizations, as when "this treatment improved average response among those in the study" is misreported as an unqualified "this treatment proved beneficial."

Given the inevitable oversimplifications of statistical methods, safe use requires treating their outputs as *guidelines* for reasoning toward inferences in ideal cases, rather than as correct answers to contextual questions. To partially address possible sensitivity of results to methods, different methods can be used in tandem with their results contrasted and merged to improved inferences and decisions (Box 1980). For example, much is made of divisions among statisticians into various schools of thought such as frequentist vs. Bayesian, each with their own toolkit of methods. Yet sound use of both requires that their methods be demoted from rules for inference or decision (as they are usually presented), and instead be treated as different viewpoints on the relation of the data to the original contextual questions, which are then integrated into an evidence synthesis or "triangulation" (Guitierrez et al. 2024).

There is a long history of advice to subsume statistical methods within informal contextual reasoning, rather than allow formal statistical inferences to dominate e.g., see Skellam (1969, p. 466), DeFinetti (1975, p. 279), and Box (1990, p. 448). Conventional methodology nonetheless sets aside these concerns to focus on development of computational methods that force out inferences based on extensive simplifying assumptions, which may look unrealistic when accurately translated into a contextual narrative. Complicating matters, many analysis choices are not specified by conventional methods and are left to the investigative team or the authorities they follow (e.g. how to set category boundaries, how to build models), leading into a "garden of forking paths" (large researcher degrees of freedom) among choices (Gelman and Loken 2014a, 2014b; McElreath 2020). These informal choices will vary considerably across studies and can largely dominate the results. Worse, the choices can easily be guided (whether consciously or unconsciously) to favor whatever result the investigators expect or want to be the case.

This research reality is largely ignored by statistical methodologies, which assume a sharply defined and detailed protocol has been set out in advance with such precision that every step could be programmed, allowing the entire analysis to be carried out by entering the raw data in a computer and reproduced to unlimited precision. Even were this done, different research teams would use different programs based on their own methodological training, and it is not clear that imposing a uniform protocol would do more than ensure the defects of the protocol go undetected.

In light of the aforementioned issues, studies have given the same dataset to different well-qualified research teams. Perhaps unsurprisingly, the results from different teams varied greatly, sometimes with conflicting results (Silberzahn et al. 2018; Kummerfeld and Jones 2023). It follows that (a) statistical presentations should not be interpreted as what the data show but rather as what the analysts would argue the data show based on their preferred approach to the data, and (b) other reasonable analysts might reach quite different conclusions using equally defensible approaches.

**The fictional worlds and oversimplifications of statistical methods**

With the above cautions in mind, we may view each statistical analysis as a *thought experiment* in a fictional "small world" or "toy example" sharply restricted by its simplifying assumptions. The questions that motivated the study must be translated properly into this fictional world; statistical methods then answer the questions via mathematical deductions from the assumptions. But those answers apply with logical force only within a fictional world in which all the assumptions hold. Thus, when key assumptions have not been perfectly enforced by design or circumstance, a crucial task is to judge how well those answers correspond to reality.

In ordinary engineering parlance, the entire set of assumptions is sometimes called a *model*, and that abbreviation will be used here. This is in conflict with its narrow usage in statistics, where a model is usually an equation representing one measured variable as a function of other measured variables and an unmeasured random-error (noise) variable. This equation summarizes many assumptions in a form that is easy to manipulate algebraically, but is so compact and abstract that some of the most important assumptions it encodes get overlooked by researchers.

The answers obtained from the model may still provide useful insights or hints about the reality targeted by the study, even when the model assumptions are inaccurate. Nonetheless, it is a common fallacy to treat model-based conclusions as if they apply directly to reality. Doing so is an example of confusing statistical fiction with reality, sometimes called *model reification*, and the usual result is severely overconfident inferences about that reality.

*Statistical models for associations: descriptive vs. causal analyses*

*Associational* or descriptive models only show how the variables are related without regard to time ordering or causality. *Longitudinal* models additionally indicate the time order of the variables but not necessarily their causal relations. *Causal models* indicate how different outcomes may follow from different interventions or decisions, and how associations are determined by causal pathways (Pearl 2009; VanderWeele 2015; Pearl et al. 2016; Hernán and Robins 2025); see **<cross-reference to the chapter on Causal Reasoning and Inference in**

**Epidemiology by Didelez and to Regression Methods for Epidemiological Analysis by Greenland (2024a) in this handbook>**).

Associational models are best justified in statistical surveys, where the goal is to make inferences about the state of a target population, rather than study causal connections among variables. An example would be a survey of cannabis use. Nonetheless, even for descriptive purposes, causal assumptions will be needed if the goal is to make inferences beyond the survey participants to the population from which the participants were drawn – albeit those assumptions are often not labeled or even recognized as causal.

The main assumption used by statistical methods for inferences beyond the sample is *random selection from the population*; this is an assumption that there is *no* cause of entry into the analysis other than a random-number generator used for selection from the population. If such random selection is not actually done, or refusal and non-response proportions are high, inferences depend on accurate and complete measurement of and adjustment for those variables that predict selection, which can be used in attempts to adjust for violations of the random-selection assumption -- that is, we may attempt to account for *causes* of selection.

Even with random selection in place, however, statistical inferences from associational models refer only to associations, not to cause and effect; any causal inferences from them must involve considerations beyond the data. While causal models that incorporate those considerations are becoming more common, most statistical analyses are still built around associational models, even when the goal is causal inference. This practice can be done safely only if the user takes care to explicate and critique the additional assumptions needed for causal inferences, such as those illustrated in causal diagrams (Greenland et al. 1999; Glymour & Greenland 2008; Pearl 2009; McElreath 2020; Hernán & Robins 2025) and other considerations discussed in the famous list of Hill (1965); see **<cross-reference to the chapter on Causal Reasoning and Inference in Epidemiology by Didelez, to Directed Acyclic Graphs by Foraita et al. and to Basic Concepts and Methods in Epidemiology by Rothman and Greenland in this handbook>**).

*A hypothetical example of statistical oversimplification*

A pressing question for medical advice and health policy asks: What are the effects of cannabis use on mental health? (Ganesh and D'Souza 2022). Among points are

1) The central interest in this question is causal, not associational.
2) The most directly relevant long-term data will be observational, since long-term randomized trials of cannabis use are absent and infeasible.
3) The causal variable (often called the "exposure" or "treatment") targeted by the question is cannabis use. That is a complex collection of possibilities, ranging from very occasional use to continual daily use; from smoking to vaping to ingestion; from low-potency to high-potency sources; from narrow to broad cannabinoid mixtures; and from high-purity to heavily adulterated sources contaminated with pesticides, fungal toxins, chemical residues, and unlisted psychoactive adulterants.
4) All aspects of cannabis use will fluctuate over time, often dramatically, and will be strongly associated with variables such as age and local laws.

In tandem, psychiatric diagnosis subsumes a complex collection of mental states and behaviors, ranging from amotivation syndromes or social withdrawal to overt cognitive impairment or psychoses.

To illustrate typical simplifications made in basic statistical analyses, Table 1 presents a hypothetical two-by-two table of counts labeled as if from a survey of self-reported cannabis usage $X$ ($X = 0$ if none, $X > 0$ if some) and a subsequent serious psychiatric diagnosis $Y$ ($Y = 0$ if not, $Y = 1$ if so). Foremost is the collapsing of usage categories and dimensions (dose, frequency, etc.) and diagnoses (depression, psychosis, etc.) into binary indicators $X$ and $Y$. Such collapsing is often done to reduce statistical fluctuations as well as simplify presentation, but can be misleading because it can hide important details and variations in the relation under study.

Analyses should explore these details using multiway tables and multiple regression (Jewell, 2004; Greenland 2008a; Hosmer et al. 2013; Harrell 2015; Agresti 2018); see **<cross-reference to the chapter on Regression Methods for Epidemiological Analysis by Greenland (2024a) in this handbook>**). A caution with these methods however is that correlation coefficients, “variance explained”, and “standardized” coefficients will confound effects of interest with irrelevant study features (Tukey 1954; Rosenthal and Rubin 1979; Greenland et al. 1986, 1991; Cox and Wermuth 1992).

It will also be advisable to analyze uncontrolled sources of statistical bias (Fox et al. 2021, Hernán and Robins 2025); see **<cross-reference to the chapter on Sensitivity Analysis and Bias Analysis by Greenland (2024b) in this handbook>**). Statistical analyses that ignore these sources, such as those discussed below, should be seen as providing only inferences about associations between the measurements produced by the processes that generated the data, which represent a blend of effects of interest with uncontrolled bias sources (Greenland 2022, 2025).

**Table 1. Reported cannabis usage per month cross classified against psychiatric diagnosis. $H_0$ is independence of usage and diagnosis.**

| | *X* category: $X = 0$ | $X > 0$ | Total |
|---|---|---|---|
| Diagnosed ($Y = 1$): | | | |
| Observed | 16 | 10 | 26 |
| Expected under $H_0$ | 20.8 | 5.2 | 26 |
| Not diagnosed ($Y = 0$): | | | |
| Observed | 464 | 110 | 574 |
| Expected under $H_0$ | 459.2 | 114.8 | 574 |
| Total | 480 | 120 | 600 |
| Proportion diagnosed | 16/480 | 10/120 | 26/600 |
| Percent diagnosed | 3.3 | 8.3 | 4.3 |
| Expected percent under $H_0$ | 4.3 | 4.3 | 4.3 |
| Risk difference *RD* | 0 | 0.050 | |
| Risk ratio *RR* | 1 | 2.5 | |
| Odds ratio *OR* | 1 | 2.6 | |

Despite the extreme contextual simplifications embodied in a two-by-two table of counts, such presentations remain common and can provide an illustration of basic concepts, methods, interpretations, and misconceptions in conventional uses of statistical methods. Sound analyses begin with simple description of the data. In Table 1 the proportion with diagnosis recorded among those reporting usage in the survey is 0.083 or 8.3%, and among those reporting no use is 0.033 or 3.3%, corresponding to diagnosis odds of 0.083/(1−0.083)=0.0909 and 0.033/(1−0.033)=0.0345. These produce a risk difference (*RD*) of 0.050 or 5%, a risk ratio (*RR*) of 0.083/0.033=2.5, and an odds ratio (*OR*) of 0.0909/0.0345 = 2.6. The proportions 0.083 and 0.033 are the mean values of the binary indicator *Y*, hence the *RD* is the difference in the mean of *Y* between the *X* groups. These measures of association represent a blending of any real usage effects with biases from selective survey participation, self-selection for usage, and misreporting or miscoding of usage and diagnosis.

By any measure, Table 1 exhibits an association of usage with risk. To address whether the association is attributable to diagnoses among heavy users, we could examine the two-by-two classification comparing lighter to heavier usage, e.g., “4 times a month or less” vs. “more than 4 times a month”. Nonetheless, all basic statistical methods assume the analysis was prespecified; that is, they assume analysis choices were made without regard to the results they produced. Choosing variables or statistics to maximize the association will produce estimates that are too large and *P*-values that are too small, while choosing variables or statistics to minimize that association will produce estimates that are too small and *P*-values that are too large. In either case, the resulting interval estimates will be too narrow, thus understating the appropriate level of uncertainty about the results.

**_P_-values, statistical tests, and interval estimates**

The theory and terminology surrounding *P*-values is quite confusing, and as a result *P*-values, “significance tests”, and their associated interval estimates (usually called “confidence intervals") may be the most misunderstood, mis-taught, and misused methods in the history of science. Surveys have indicated that most users cannot correctly define or interpret any of these concepts, and that many primers give incorrect descriptions (Haller and Krauss 2002; Lecoutre et al. 2003; Greenland et al. 2016b). Even writings that give technically correct definitions usually fail to emphasize how these methods depend on assumptions that are often implausible in research on human populations. Therefore, the present chapter provides a framework and language that encourages their proper use and interpretation.

*Selection into surveys*

Consider first a survey intended to produce inferences about the features of a larger, sampled *target* population, that will measure a set of covariates on each person in the survey. The central assumption of statistical survey methods is that selection is *conditionally random* (or some functionally equivalent assumption): for every person in the population, their chance of entering into the final analysis is known or can be unbiasedly estimated (at least up to a constant of proportionality). This assumption refers both to a person's selection for study and to subsequent events that lead to the inclusion of the person's data in the analysis. Thus, denoting the analysis-selection indicator by $S$ ($S = 1$ if selected, 0 if not), in the cannabis example $S = 1$ only for the final 600 persons in the analysis. If the goal was to estimate cannabis usage beyond those who completed the survey, $S$ would be 0 for those who were either not invited or invited but not included in the analysis because they did not supply usable data.

Suppose those invited were a random sample of size 1,000 from a Health Maintenance Organization (HMO) population. It would seem highly implausible that the final 600 in the analysis are a random sample of the 1,000 invited or of the entire target population, or that the chance of inclusion in the analysis can be unbiasedly estimated from the covariates available in the HMO database. Ordinary statistical inferences about cannabis usage would therefore be restricted to persons in the HMO population who would have supplied usable data had they been invited. This restriction is sometimes summarized by saying that inferences are limited to persons who are *exchangeable* with persons in the analysis data. The hypothetical collection of such exchangeable persons is sometimes called the *source population* for the data. Unfortunately, we can only delineate the form and relation of this population to the original target population if the covariates we measured can accurately predict participation.

*Experimental studies of causation*

Consider next an experiment intended to produce inferences about the effect of different treatment choices on subsequent health outcomes. The experimenters must first recruit a study population composed of those who consent to be assigned their treatment by the experimenter and have their outcomes and relevant covariates obtained by the experimenters. Because the recruited population will be highly selective (e.g., due to age or health restrictions), some methodologists treat this recruited population as the target for causal inferences (Hernán and Robins 2025).

The central assumption of statistical methods for estimating treatment effects is that treatment assignment is *conditionally random* or some functionally equivalent assumption: For everyone in the experiment, the chance of being assigned a particular treatment is known or can be unbiasedly estimated from what is known (e.g., via regression on measured pre-treatment covariates). This requirement is sometimes referred to as conditional exchangeability or conditional ignorability of treatment assignment, no uncontrolled confounding, or "no unmeasured confounding" (which is often mistaken for but is not the same as no confounding; see Greenland and Robins 2009; Greenland and Mansournia 2015; Deaton and Cartwright 2018).

Even if assignment is conditionally random, however, ordinary statistical inferences about treatment effects beyond the experimental population are limited to persons who would have met all conditions necessary for recruitment and retention, including identification and consent (Deaton and Cartwright 2018). This limitation is again summarized by saying that statistical inferences are restricted to persons who are *exchangeable* with persons in the analysis data. Recruitment restrictions are imposed by the experimenter based on measured covariates (such as age and health history), and we can identify persons meeting these restrictions. Unfortunately, as in the survey case, we can only fully identify the exchangeable population if the covariates we measured can accurately predict participation. Furthermore, the composition of this population, and thus treatment effects in it, are likely to differ substantially from populations of practical interest. In modern literature, this problem of generalizing causal inferences from experiments is dealt with under the topic of *transportability* (Dahabreh et al. 2024).

There is a logical parallel between the randomness assumptions of surveys and experiments which allows certain methods for survey analysis to be applied to experiments. We can think of each treatment group as a sample from the recruited experimental population, where each sample will be subjected to treatment different from the others. In this sense, each treatment group is a sample from a small enumerated population composed of those recruited, and if treatment is randomized each group will be a random sample from that population. The difference from a survey is that in the experiment every population member will be in one of the samples and will be assigned a treatment.

As with surveys, experiments often undergo some post-recruitment attrition in which persons may withdraw or be lost to follow-up in a manner that leaves their outcome unknown, in which case their outcome is said to be *censored*. Others may be retained but stop their assigned treatment (e.g., because it is burdensome or has unpleasant side effects), in which case they are said to be *non-adherent*. Standard methods may retain these persons in the analysis but assume *conditionally random* censoring: For every person in the analysis, their chance of censoring or non-adherence is known or can be unbiasedly estimated (at least up to a constant of proportionality) from what is known about them, i.e., the covariates. The assumptions and analysis details for these methods can be quite involved (Hernán & Robins 2025).

*Non-experimental (observational) studies of causation*

For many important questions, experiments on human subjects are impractical, unethical, or both. For example, concerning effects of cannabis use on health, few people would submit to assigned usage levels for extended periods of time, and such an experiment would not be approved. Yet the need for study seems indisputable.

A pressing question is then: What can we learn from purely observational studies? The answer depends crucially on the study design, the accuracy of measurements, and the method of analysis, which determine how accurately we can predict targeted exposures and outcomes from the measured covariates. Many specialized methods have been developed to address these issues based on modeling factors that could affect observed associations and thus distort estimates of parameters targeted for study (e.g., Greenland 2005; Carroll et al. 2006; Gustafson 2015; Little and Rubin 2019; Fox et al. 2021;  Hernán and Robins 2025); see **<cross-reference to the chapter on Causal Reasoning and Inference in Epidemiology by Didelez, to Sensitivity and Bias Analysis by Greenland, to Regression Methods for Epidemiological Analysis by Greenland, to Confounding, Interaction, and Effect-Measure Modification by Pearce and Greenland, and to Misclassification by Gustafson and Greenland in this handbook>**). These bias sources include confounders, non-random selection for analysis, non-randomly missing data, measurement and classification errors, data coarsening, data sparsity, and incorrect adjustments.

Identification of bias sources is central to valid analysis and interpretation, especially when there was no randomization and the available data are insufficient to adjust for that problem. *Causal diagrams* can be especially helpful for recognizing bias sources (Greenland et al. 1999; Glymour and Greenland 2008; Pearl 2009; Hernán and Cole 2009; Hernán and Robins 2025; see **<cross-reference to the chapter on Directed Acyclic Graphs by Foraita et al. in this handbook>**)), while magnitudes of uncontrollable biases can be addressed via analysis of sensitivity of inferences to assumptions, or *bias analysis* (Fox et al. 2021, VanderWeele 2015); see **<cross-reference to the chapter on Sensitivity Analysis and Bias Analysis by Greenland (2024b) in this handbook>**).

For Table 1 we should imagine the diverse factors that might have contributed to the apparent association beyond the effect of usage on psychiatric outcomes. For example, we should anticipate that pre-survey medical conditions (including psychiatric ones) affected both usage and psychiatric outcomes; as a result, some if not all of the association might be due to confounding (mixing) of effects of pre-existing conditions with cannabis effects.

Other contributing confounders could include age, usage of other drugs (whether recreational or medical) and alcohol. Confounding effects of accurately measured confounders can be blocked or removed (adjusted for) by a vast array of "confounder control" methods that involve matching, stratification, and statistical modeling (Lash et al. 2021; Hernán and Robins 2025); see **<cross-reference to the chapter on Causal Reasoning and Inference in Epidemiology by Didelez, to Regression Methods for Epidemiological Analysis by Greenland (2024a), and to Confounding, Interaction, and Effect-Measure Modification by Pearce and Greenland in this handbook>**). Nonetheless, while accurate measurement of age is simple, it can be difficult to impossible for recreational drugs and alcohol, and confounder measurement errors will degrade methods for confounding control. Unmeasured confounders present an even more obstinate problem, although the confounding they produce may be reduced somewhat via their relation to covariates that are controlled.

Association of usage with both non-participation (whether initial refusal or subsequent failure to supply information) and psychiatric diagnosis is another bias source. Some of the usage association with non-participation may be due to a covariate that affects them both (e.g., age); this type of selection bias parallels confounding, in that control of that covariate will then remove the component of the association due to the covariate's effects. Unfortunately, the selection bias produced from sources like cannabis effects on participation may be irremediable.

Effects of measurement errors in the primary study variables (here, usage and outcomes) can be complex and extremely difficult to control and analyze. A very old heuristic is that, if these errors are purely random, their impact will usually be to attenuate the observed association. Unfortunately, this heuristic fails when errors in variables are associated with each other and with the underlying variables (as is usually the case for questionnaire items). For example, errors in reported cannabis usage are likely to be associated with underlying psychiatric states, so this heuristic would seem unjustified in the example.

A further complication is the coarsening in Table 1: $X$ is not a treatment indicator because it does not specify how usage is determined or distributed within the very broad $X > 0$ category, yet the association will be sensitive to this distribution. Thus, we should not expect this association to apply to another population, for the latter may have a very different distribution of usage among users. Furthermore, even if cannabis effects contributed to the association seen in Table 1, we cannot say those are effects of use vs. non-use except perhaps specific to the precise distribution of use among study participants. In the causal-inference literature, this problem is taken up under the topic of *causal consistency* (VanderWeele 2015; Pearl et al. 2016; Hernán and Robins 2025); see **<cross-reference to the chapter on Causal Reasoning and Inference in Epidemiology by Didelez in this handbook>**).

**Basic theory for *P*-values and their affiliated estimates and tests**

The concept of a *P*-value can be traced to the early 1700s and became a common statistical device in the early 1800s, albeit not by that name, while the notion of "statistically significant" results appeared by the 1880s (Stigler 1986; Shafer 2020). The terms "value of *P*" and "*P*-value" appeared by the early 20th century, with the term "significance level" also being used for the same concept (Shafer 2020; Greenland 2023a).

In the form found in Pearson (1900) and Fisher (1934), a *P*-value is an index of compatibility, consonance, or consistency of the data with a model for or hypothesis about the behavior of the processes that generated the data, such as model in which cannabis use increases risk of psychiatric problems. It can also be interpreted as indicating the goodness of fit of the model to the data (Pearson 1900). This type of *P*-value is scaled from 0 = completely incompatible or complete misfit to 1 = perfectly compatible or perfect fit.

In this compatibility usage it is very important to note that high or even perfect compatibility or fit does *not* imply that the data support the model or hypothesis, because data are always highly compatible with an infinitude of other models for their generation. Conversely, low compatibility does not by itself imply that the hypothesis or model is incorrect. For example, it is always a physical possibility that the data have been altered or even fabricated to appear to have low or high compatibility with a hypothesis or model, as is sometimes discovered via demands for the data (Rubenstein 2009). Such incidents show an assumption of trust in the data underlies most applications of statistical methods – although the methods can sometimes be modified to check compatibility with the hypothesis of data integrity (e.g., Fisher 1936, p. 129-132) – see below.

Even without such problems, *P*-values and their related statistics are frequently misinterpreted to represent or imply inferences that they do not support, such as probabilities of hypotheses (Greenland et al. 2016b; Wasserstein and Lazar 2016). Thus, the following sections will explain in detail what these statistics actually represent.

*Models, statistical hypotheses, and auxiliary assumptions*

Correct understanding and use of *P*-values requires a degree of abstract theory unfamiliar to most users. To outline the original theory, define the analysis model *M* as the set of assumptions about the behavior of the process that generated the data. Those assumptions may include a *statistical hypothesis H* that is targeted for study; this hypothesis is also called the *target hypothesis* or *test hypothesis*. An example in the cannabis study is the *null hypothesis* that reported usage *X* and psychiatric outcome *Y* have no association in the source population, which can be equivalently expressed as $RD_{\text{pop}} = 0$, $RR_{\text{pop}} = 1$, or $OR_{\text{pop}} = 1$. As another example, *H* could be that, for every monthly usage frequency *f*, persons reporting frequency $f + 1$ exhibit a 4% higher odds of a psychiatric diagnosis ($Y = 1$) than those reporting usage *x*.

It should be noted that many if not most statistical writings use the term "null hypothesis" for any statistical hypothesis. This usage is however at odds with ordinary English, in which "null" means nothing or zero (Greenland 2023a, Sec. 5.5); hence other writings and the present chapter use instead "target hypothesis" and reserve the term "null hypothesis" for hypotheses that assert no association or no effect. In this usage, examples of non-null target hypotheses include:

- the odds of diagnosis among users is twice that among non-users ($OR_{\text{pop}} = 2$);

- the odds among users is within half to twice that of non-users ($\frac{1}{2} \le OR_{pop} \le 2$);
- the odds among users is no higher than that of non-users ($OR_{pop} \le 1$); and
- the odds among users is no more than 20% higher than that of non-users ($OR_{pop} \le 1.2$).

The hypotheses $OR_{pop} = 1$, $OR_{pop} = 2$, and $\frac{1}{2} \le OR_{pop} \le 2$ are examples of two-sided hypotheses, whereas $OR_{pop} \le 1$ and $OR_{pop} \le 1.2$ are examples of one-sided (or directional) hypotheses. The hypotheses $OR_{pop} = 1$ and $OR_{pop} = 2$ refer to only one possible value for the target parameter and are usually called *point hypotheses*. Unless noted otherwise, the remainder of this chapter will assume *H* is a point hypothesis.

Besides the target hypothesis *H*, the model will also include crucial auxiliary assumptions that are used to compute *P*-values and estimates; the entire set of these assumptions will be denoted by *A*. One common auxiliary assumption is that the final data are a stratified random sample from the target population to which *H* applies; this assumption would be incorrect if the variables targeted (above, cannabis use and diagnosis) both affect consent to participate or inclusion in the final analysis data.

Another common auxiliary assumption is that outcomes are independent across persons; this assumption would be incorrect if the persons in the study interact ("interfere") with each other in ways that affect each other's outcomes (as when the diagnosis of one person affects the diagnosis of other persons). In that case, *P*-values computed using the no-interference assumption would be invalid.

Another auxiliary assumption used in regression analyses would be that outcome trends in the source population follow the shape specified by the regression model; see **<cross-reference to the chapter on Regression Methods for Epidemiological Analysis by Greenland (2024a) in this handbook>.** For example, a logistic shape is usually assumed to compute *P*-values for trends in odds, such as the *H* for use frequency *f* mentioned above.

*Discrepancy statistics and reference distributions*

The full analysis model *M* is the combination of the target hypothesis *H* and auxiliary assumptions *A*, and is denoted $M = H$ & $A$ or $M = (H, A)$. This full model is used to make specific predictions about what we should expect in the data if the model is correct. From those predictions and the data, the model is then used to derive a *statistic T* (a formula to be applied to the data) that summarizes the discrepancy or divergence between the data and the model predictions. The model is then used to derive a *reference distribution* or *sampling distribution* for *T* that shows how we would expect *T* to vary across random samples if *M* were correct, at least to an approximation that improves as the sample size increases. *T* is traditionally called a "test statistic", but we avoid that usage here because *T* and its *P*-value can be used for estimation of association and effect measures.

There may be many ways to summarize the divergence of the data from the model predictions. The statistic *T* is usually chosen to be most sensitive to (or "most powerful for") violations of *H* when *A* is taken as correct or "given", in the sense that discrepancies resulting from those violations will inflate *T* to a maximal extent. Unfortunately, choosing *T* based solely on its sensitivity to *H* violations can have adverse consequences when *A* contains false assumptions;

for example, if $T$ is chosen to be most sensitive to linear trends, it will have poor sensitivity to U-shaped trends.

Examples of $T$ include $Z$-statistics, $\chi^2$, $F$, Student-$t$ statistics, and many other formulas for measuring discrepancy between the data and what $M$ predicted; their names are taken from their conventional reference distributions. As an example, suppose $T$ is the familiar Pearson $\chi^2$ statistic for evaluating the hypothesis $H_0$ that the $J$ row and $K$ column classifications of a two-way table are independent in the source population, where the subscript zero on $H_0$ denotes "no association". The formula for $T$ is then $\Sigma(O - E)^2/E$, where the $O$ is an observed cell count, $E$ is the count expected (predicted) from $M = (H_0, A)$, and the observed marginal totals, and $\Sigma$ denotes summation over all the cells. If in addition to assuming $H_0$ we make the auxiliary assumptions ($A$) that there is no interference between outcomes of persons and there are no problems of data integrity, this $\chi^2$ statistic has an approximate reference distribution that is $\chi^2$ with $(J-1)(K-1)$ degrees of freedom (Agresti 2018).

In the cannabis example, the row and columns classifications are diagnosis and cannabis usage, the sum is over 4 cells, and $H_0$ says that usage and diagnosis are not associated (i.e., $OR_{\text{pop}} = 1$). The cells expected under $H_0$ are given in the Table 1; they are the products of the row and column margins divided by the total of 600. From those, the Pearson $\chi^2$-formula yields a value of 5.79, and the approximate reference distribution is a $\chi^2$ distribution with $(2-1)(2-1) = 1$ degree of freedom. The same general Pearson formula can be used for non-null hypotheses, e.g., that $OR_{\text{pop}} = 2$, although then computation of the expected values $E$ becomes more involved.

Now suppose we have a point hypothesis *H*, a statistical formula *T* that measures the discrepancy of the data from what *H* leads us to expect, and the observed value *t* for *T* computed from the data. Then the reference distribution gives an approximate probability *p* that *T* would be as or more extreme than the observed *t* if all assumptions used to compute the distribution (including both *H* and *A*) were correct. This is the traditional "tail-probability" definition of a *P*-value. When "extreme" corresponds to *extremely large T*, this tail-probability *p* can be defined abstractly as the probability that *T* would have fallen at or above the observed value *t*, given the assumptions encoded in the model $M = (H, A)$; in mathematical terms, $p = \Pr(T \geq t \mid M) = \Pr(T \geq t \mid H, A)$. Thus, referring the statistic of 5.79 from Table 1 to a $\chi^2$ reference distribution with 1 degree of freedom, we obtain $p = \Pr(T \geq t \mid M) = 0.016$.

In all these cases, the tail-probability *P*-value *p* is the observed "value of *P*" or "significance level" as defined by Pearson (1900) and Fisher (1934, 1936), which in the 1920s came to be called the *P*-value, and is how *P*-values are defined in most of applied statistics. Unfortunately, some mathematical-statistics texts redefine *P*-values based on decision rules for hypothesis testing (discussed below). In practice, for point hypotheses the two definitions usually lead to numerically identical *P*-values; however, in other cases such as for interval hypotheses (e.g., $½ \leq OR_{pop} \leq 2$) a tail-probability *P*-value can be larger than a decision *P*-value (Greenland 2023a).

It should be noted that there are situations in which the model implies that very small values for *T* are improbable (as with a $\chi^2$ statistic with 3 or more degrees of freedom) and "extreme" corresponds to *extremely small T*. This situation arises when there are concerns that the model fits the data *too* well, suggesting data manipulation (Fisher 1936, p. 129-132); in these cases the *P*-value becomes the lower-tail probability $p = \Pr(T \leq t \mid M)$.

*Approximate vs. exact reference distributions*

Beyond basic data structures, such as 2-by-2 tables, finding an exact reference distribution for $T$ and then computing a *P*-value from it can be difficult. Thus, conventional reference distributions are often "large-sample" approximations, and the resulting *P*-value becomes inaccurate as $t$ becomes more extreme or the sample size becomes small. For example, the Pearson $\chi^2$ statistic $T = \Sigma(O - E)^2/E$ is named after its conventional reference distribution, the $\chi^2$-probability distribution. This approximation is based on assuming each observed count would vary across random samples following a normal distribution with mean $E$ and variance $E$. If the model is correct, the approximation becomes more accurate as the expected values $E$ increase. But, because the $O$ are counts, they cannot be exactly normal and so the $\chi^2$-distribution provides only an approximate *P*-value for the Pearson statistic.

Again, for Table 1 the $\chi^2$-reference distribution has 1 degree of freedom, and the observed approximate *P*-value is $p = 0.016$. Thus the probability $\Pr(T \geq 5.79 \mid H, A)$ that $T$ would fall at or above $t = 5.79$, *if* indeed $H$ and $A$ were correct, is approximately 1.6%. In 2-by-2 tables, however, the exact reference distribution for $T$ (called the *hypergeometric*) produces instead Fisher's exact *P*-value (Agresti 2018) of 0.041, or 4.1%. It is possible to adjust approximations such as the Pearson *P*-value to come closer to exact *P*-values; a more general alternative enabled by modern computation is to use simulation methods such as the bootstrap (Harrell 2015; Agresti 2018).

*Interpreting P-values via percentiles and surprisals*

Correct interpretations may be aided by appreciating the asymmetry inherent in the information *P*-values provide. A *P*-value is *not* a probability of the hypothesis *H* used to derive it. Instead, it is only the quantile or percentile at which a statistic *T* fell in a reference distribution, where the distribution is used to gauge how unusual or surprising *T* would be *if H and all the auxiliary assumptions used to derive T were correct*. Thus, using the statistic *t* to measure divergence, 100*p* shows how far the data diverge from the model predictions in percentile terms; for example, 100*p* = 4% shows that *t* just reached the top 4% of possible divergences.

An analogy would be the percentile at which a person taking an examination fell in the reference distribution of everyone else taking the exam. If we had thought in advance that this person had at most an average understanding of the exam topic relative to typical examinees, we might be surprised if the upper percentile location of the person's exam score was 4.1%, placing their score in the top 5% of the score distribution. Similarly, if we had thought in advance that cannabis usage and psychiatric outcome would *not* be associated (as per $H_0$) among survey responders, we might be surprised if the exact statistic from Table 1 fell at the upper 4.1 percentile of its reference distribution, in the top 5% of that distribution.

We may make the degree of surprise in a *P*-value more tangible if we imagine a fair coin-tossing mechanism and then ask how surprised we would be if the mechanism gave us all heads in a test comprising *n* tosses, with *n* is chosen so that the probability of *n* heads in n tosses ($½^n$) is closest to our observed *P*-value *p*. For *p* = 0.041, this *n* is 5, because $½^5$ = 0.031 and $½^4$ = 0.063. We could then say that our exact *P*-value from Table 1 was a little less surprising than seeing 5 heads in exactly 5 fair tosses.

More generally, given a *P*-value *p*, we look for the *n* closest to the binary *S-value* $s = -\log_2(p)$, also known as the *surprisal* or *Shannon information* against the model $M = (H, A)$ provided by the *P*-value *p*, where *p* is the percentile at which *T* fell in its reference distribution under the model (Rafi and Greenland 2020; Cole et al. 2021; Amrhein and Greenland 2022; Greenland et al. 2022; Greenland 2023a). This *S*-value can also be used as a measure of incompatibility between the data and the model. The units of binary surprisal are *bits* of information, as found in computer science and electronics, and its scale runs from 0 (representing no information conveyed by *T* against the model) upward without bound.

The information in one bit is very small, equivalent to the information about fairness conveyed by one coin toss. Consider again a coin being tossed *n* times to determine whether the tosses are biased toward coming up heads, where now *n* is the integer nearest *s*. The *S*-value *s* can then be interpreted as a measure of the information against the hypothesis of no bias for heads conveyed upon observing all heads in the *n* tosses. For Table 1, $s = -\log_2(0.041) = 4.6$; thus, seeing $t = 5.79$ supplies 4.6 bits of information against independence $H_0$ ($OR_{\text{pop}} = 1$), given the auxiliary assumptions *A*. This result conveys slightly less information against $OR_{\text{pop}} = 1$ than seeing 5 heads in 5 tosses conveys against there being no bias toward heads. For comparison, the *S*-value for seeing $p = 0.05$ is 4.3, hardly more information against the tested hypothesis than seeing 4 heads in 4 tosses provides against no bias toward heads.

Many researchers would *not* have expected there to be no association of cannabis and psychiatric outcomes, given the many ways we can imagine they might be associated; thus, the null hypothesis $H_0$ is a "straw man" and its *P*-value of 0.041 should not be seen as surprising given actual background (contextual) expectations. We can however compute *P*-values for more contextually reasonable hypotheses (Greenland 2017; Rafi and Greenland 2020; Amrhein and Greenland 2022; Greenland et al. 2022). For example, suppose we had expected to see a doubling of the odds of diagnosis among users compared to non-users. The corresponding statistical hypothesis *H* is that the odds ratio $OR_{\text{pop}}$ in the source population for Table 1 equals 2 ($OR_{\text{pop}} = 2$); the exact *P*-value for this hypothesis is $p = 0.644$. This result is unsurprising contextually and statistically: $s = -\log_2(0.644) = 0.6$ bits, less than the information about fairness contained in one coin toss, showing Table 1 contains negligible information against $OR_{\text{pop}} = 2$.

*The narrowness of statistical information*

It should be noted that the information measured by *T*, and thus by its *P*-value and *S*-value, is very narrow. *T* represents only one dimension of possible discrepancy of the data from what the full model predicts, such as the divergence of observed outcomes from expected or predicted values. Many other dimensions can and often should be checked, such as observed dispersion of outcomes compared to predicted dispersion. For quantitative outcomes, underdispersion refers to outcome values appearing more clustered than predicted, and may indicate the presence of interference (such as contagion); while overdispersion refers to outcome values appearing more spread out than predicted, and may indicate the presence of important uncontrolled outcome predictors.

Furthermore, *T* and its *P*-value and *S*-value summarize only deductive (purely logical) information about the relation of the data to the full model *M*, and their computation makes no use of contextual (background) information beyond that captured in *M*. In particular, if *M* contains no causal assertion, the statistics derived from it summarize only information about associations; conversely, for statistics to summarize causal information, the full model *M* must contain causal statements (Pearl 2009; Hernán and Robins 2025). For example, the statistics from Table 1 make no use of the contextual information that users and non-users differ in many ways that could affect the outcome (i.e., differences in potential confounding factors). We should thus see $p = 0.041$ as referring only to the hypothesis $H_0$ in *M* that cannabis use and outcome are *unassociated* in the source population.

If instead $H_0$ is the causal null hypothesis that cannabis use has *no effect* on outcome, asserting that $p = 0.041$ refers to $H_0$ would require the auxiliary assumptions (*A*) include that cannabis use had been randomized, which is contextually absurd. A more sensible interpretation would thus restrict $p = 0.041$ to refer only to "no association", not "no effect".

*Interval estimation and its relation to P-values*

To avoid misinterpretations, we can compute *P*-values for multiple values of an association measure such as $OR_{\text{pop}}$ and then tabulate or plot them against the association values to create a *P*-value (or compatibility) function (Poole 1987; Rafi and Greenland 2020; Amrhein and Greenland 2022); similarly, we can plot the resulting *S*-values against the association values to create an *S*-value (or surprisal) function (Rafi and Greenland 2020). Although such a table or graph would be highly informative, it can be too bulky for reports involving more than one association. Thus, the usual convention presents association values that have pre-specified *P*-values. One is the point estimate, which is the value of the association at which the *P*-value is largest (usually $p = 1$); in Table 1 that is $OR_{pop} = 2.64$. The others are the association values for which the *P*-value equals a small number $\alpha$, usually 0.05, which may be called $\alpha$-level *compatibility limits* (CL) and correspond to what are usually labeled as $100(1-\alpha)\%$ "confidence limits". The association values between these limits are those for which $p > \alpha$. For Table 1, the exact *P*-value is 0.05 for odds ratios of 1.04 and 6.36; thus, these two odds ratios are the exact 0.05-level CL.

The range within the CL is called a *compatibility interval* (CI) (Rafi and Greenland 2020; Amrhein and Greenland 2022; Greenland et al. 2022; Rovetta et al. 2025). Association values within a CI have $p > \alpha$ and thus could be said to be "reasonably compatible" with the data if the criterion for that is having $p > \alpha$ (given all the auxiliary assumptions $A$). In Table 1, the range between 1.04 and 6.36 is a 0.05-level CI for $OR_{pop}$; thus, we can say that all values for $OR_{pop}$ between 1.04 and 6.36 are reasonably compatible with the data, under the auxiliary assumptions and using $p > 0.05$ as our criterion for "reasonable compatibility".

*Coverage ("confidence") intervals*

The interval between 1.04 and 6.36 is more familiarly labeled as a $100(1-0.05)\% = 95\%$ confidence interval, a term avoided here because it implies there are grounds for confidence, and those grounds may be uncertain or false (Amrhein et al. 2019a,b; Rafi and Greenland 2020; Amrhein and Greenland 2022; Greenland et al. 2022). Even when those grounds are satisfied, however, the "confidence" corresponds only to assurances about the coverage rate (or calibration) of intervals generated in a thought experiment that assumes ideal study conditions.

More precisely, under the auxiliary assumptions $A$, we can imagine the observed data are a random sample from the source population, where the specific meaning of "random" depends on the study design. We may then ask what proportion of all possible random samples from the source population produce $P$-values and intervals with certain properties, given the auxiliary assumptions in $A$ are correct. For example, we may ask: Given the assumptions, in what proportion of these samples will the $\alpha$-level CI contain or *cover* the odds ratio $OR_{\text{pop}}$ computed from the entire source population? Since the interval contains the odds ratios for which $p > \alpha$, this coverage question is the same as asking: In what proportion of random samples will the $P$-value for $OR_{\text{pop}}$ exceed $\alpha$ if $A$ is correct?

The actual proportion of samples in which coverage occurs is called the *coverage rate of the method for computing the intervals*; the proportion for which coverage fails is called the *error rate of the method*. If the coverage rate is no less than $1-\alpha$ when all the auxiliary assumptions are satisfied (as with exact intervals), the method is said to be *conservatively calibrated* or have *conservative coverage validity* under the assumptions. The "conservative" part of the terms arises because the criterion for validity is coverage no less than $1-\alpha$ rather than equal to $1-\alpha$; conservative validity at the 95% level means the coverage is *at least* 95%.

The definitions of coverage and error rates refer to what the method produces across all possible random samples under ideal conditions, rather than to the observed interval. For an observed interval, such as 1.04 to 6.36, all that can be said is that it was produced by a method designed to ensure conservative coverage under ideal conditions; that is, *if all the auxiliary assumptions are* *correct*, at least 95% of the intervals computed by the method from all possible random samples should contain $OR_{\text{pop}}$. The coverage gets closer to 95% as the sample size increases, but its difference from 95% can be appreciable when samples are small or sparse.

Unfortunately, tradition has led to observed compatibility intervals being called "confidence intervals" and $1-\alpha$ being called the "confidence level" of $\alpha$-level compatibility intervals. These terms can be highly misleading because the "confidence" only refers to how the method performs across all possible samples *under the assumptions used to compute the intervals*. Thus, in the example the 95% refers to coverage rate of the *method* only; the 95% does *not* refer to the probability that the observed interval (1.04, 6.36) contains $OR_{\text{pop}}$. For that specific interval (1.04, 6.36) to have a probability of containing $OR_{\text{pop}}$, $OR_{\text{pop}}$ must be given a prior distribution (as in Bayesian methods); when that is done, the probability that (1.04, 6.36) contains $OR_{\text{pop}}$ may not be near 95%. The general rule is thus that one should not confuse the "confidence" in "confidence interval" with a probability that the observed interval contains the true parameter $OR_{\text{pop}}$.

A related problem is that, when some of the assumptions are of uncertain status, $1-\alpha$ can vastly overstate the confidence we should have in the method; in other words, the CI becomes an *overconfidence* interval. Thus, in the example the coverage rate could be much less than 95% if some assumptions were violated, e.g., if the diagnoses were not independent (as might be suspected if they were all made by the same psychiatrist), or if survey participation was associated with both usage and diagnosis. In contrast, even if the assumptions are uncertain or thought to be false, the compatibility interpretation remains valid because it refers only to how compatible the data are with hypothesized associations under the assumptions; it does not refer to whether the interval is a reliable guide to the targeted association or whether we can be confident that $OR_{pop}$ is in the interval (Amrhein et al. 2901a, 2019b; Rafi and Greenland 2020; Amrhein and Greenland 2022; Greenland et al. 2022). Such reliability judgements should instead be formed based on a detailed overview of the design, conduct, and analysis of the study as well as other research on the topic. When the presenters have potential conflicts of interest, readers might take note of those conflicts in forming judgements and should be aware that conflicts may go unmentioned despite disclosure requirements (Greenland 2009).

*Bayesian statistics*

The methods described thus far are sometimes labeled "frequentist" because the only probabilities they provide are for hypothetical event frequencies (e.g., an interval coverage rate expected under certain assumptions). *Bayesian* methods go further to create *observed* intervals that are supposed to have 95% probability of containing the targeted association (Greenland 2008b; Gelman et al. 2013; Gustafson 2015; McElreath 2020): see **<cross-reference to the chapter on Bayesian Methods in Epidemiology by Held in this handbook>**). There are many primers and books on Bayesian methods; the reader is however warned that there are important differences among philosophies and methods labeled "Bayesian" (Good 1983). Consequently, writings on Bayesian methods can differ substantially on fundamentals, and they often omit critical assessments and limitations of the fundamentals they espouse (Gigerenzer and Marewski 2015).

Unfortunately, Bayesian probability claims depend on the same auxiliary assumptions and thus can be as doubtful as the coverage claims made by "confidence-interval" methods. Furthermore, Bayesian results depend on the addition of what are known as *prior distributions* (or "priors") that represent information beyond that used to obtain the conventional frequentist results. The justification for Bayesian methods is that valid prior information can improve the accuracy of results. Nonetheless, this potential for benefit brings with it a potential for harm: Use of invalid prior information can reduce the accuracy of results. Use of priors also introduces a danger of circular reasoning: Bayesian results can be biased towards desired or foregone conclusions by using a prior distribution that favors those conclusions. It is therefore strongly advised that Bayesian results should always be preceded by conventional frequentist results, so that the reader can see the influence of the added prior distribution on the results (Greenland 2025).

Another device for judging a prior distribution is to convert it into *prior data*, that is, data that express the information in the prior as if they were from a perfect random survey or perfect randomized experiment (Bedrick et al. 1996; Greenland 2008b; Clark 2025). To illustrate, some studies have used prior distributions for a rate ratio that assert with at least 95% probability that the true rate ratio is between 1/1.20 = 0.83 and 1.20. For a simple balanced randomized trial to produce a 95% CL of 0.83 to 1.20 would require observing 232 cases of the outcome in each treatment arm, or 464 cases total (Sullivan and Greenland 2013, p. 311). Thus, a prior distribution that places 95% or more certainty on the rate ratio falling between 0.83 and 1.20 represents far more information than available from most randomized trials, and would dominate the Bayesian results from those trials – that is, the prior distribution would overwhelm the data information.

A strength of a data representation for prior information is that it allows one to see how strong the prior is in terms of a corresponding study size. It also allows one to develop and use priors that are less dominating and presumably more cautious about what has actually been observed before the present study (Greenland 2008b). A further advantage is that one can obtain a Bayesian extension of a frequentist analysis simply by concatenating a prior-data set onto the actual-data set and adding an indicator variable to the data to distinguish the actual data from the prior data. Sullivan and Greenland (2013) give details, with SAS code for Bayesian logistic, Poisson, and proportional-hazards regression; see Discacciati et al. (2015) and Greenland et al. (2016b, appendix) for Bayesian logistic regression via data priors in Stata and in R, respectively.

*Approximations in statistical software*

Most statistical software computes *P*-values and interval estimates from normal (Gaussian) approximations. To justify coverage ("confidence") and error rate ("significance") these programs require additional assumptions beyond random sampling or treatment randomization:

1) The form of the assumed statistical model is correct; and
2) The size of the data set is large enough to make the approximations accurate enough for practical purposes.

Both assumptions are often violated to an important extent, sometimes resulting in dramatic bias (Greenland et al. 2016a), yet they are ignored in the most research reports.

Perhaps the most common approximation is the *Z*-score or Wald method. Consider a point estimate $b$ of a regression coefficient $\beta$. The Wald method uses an estimate of the standard deviation of $b$ across all possible random samples that the study design could have generated. This estimated standard deviation is usually labelled the *standard error* (*SE*) of $b$ in software output. Suppose that $H$ is the hypothesis that $\beta$ equals a specific value $c$; the approximate two-sided $P$-value for $H$: $\beta = c$ is computed by looking up the $Z$-score $|b - c|/SE$ in a normal distribution table or function. The corresponding Wald 95% interval estimate for $\beta$ is the range of all values $c$ for which $p > 0.05$; its endpoints (limits) are $b \pm 1.96 \cdot SE$. This approximation is often refined by replacing the normal with a $t$-distribution, thus increasing the $SE$ multiplier above 1.96; while this refinement can be an improvement, this still is an approximation unless the regression outcome is perfectly normal.

Especially for categorical outcomes, Wald approximations become increasingly inaccurate as the sample size gets smaller or the $Z$-score becomes more extreme (Agresti 2018). At typical sample sizes, approximate $P$-values near 0.05 can be off by 10%, while $P$-values below 0.001 can be off by an order of magnitude relative to more accurate methods. In tandem, the odds-ratio estimates become increasingly biased away from 1 as the data become more sparse (signaled by extreme estimates) and is accompanied by severe inaccuracies in the Wald limits (Greenland et al. 2016a; Greenland 2021a).

**A critical overview of statistical significance tests**

$P$-values are versatile statistical tools that can provide both estimates and compatibility measures. Unfortunately, these applications have been neglected, relegating $P$-values to use only as tools for statistical tests. In dichotomous $\alpha$-level hypothesis testing, the observed $P$-value $p$ for a hypothesis $H$ is compared to a "small" pre-specified number $\alpha$; $H$ is then "rejected" if $p \leq \alpha$, and "accepted" if $p > \alpha$, so that "reject if $p \leq \alpha$" functions as a decision rule about when to abandon or keep *H*. To avoid misinterpretations, many sources replace "acceptance" with "fail to reject", so the $\alpha$-level rule becomes "reject *H* if $p \leq \alpha$; don't reject if $p > \alpha$".

Rejection of a correct *H* is known as a "false-rejection", "false-positive", "Type-I", or $\alpha$ error; failing to reject an incorrect *H* is known as a "false-acceptance", "false-negative", "Type-II", or $\beta$ error. These notions of error should be contrasted to the continuous notion of error in estimation, in which the degree of error is the distance from the estimate to the correct value of the targeted association or effect.

Often, the test (decision rule) is implemented using CI instead of *P*-values, as "reject *H* if the $1-\alpha$ interval excludes the association specified by *H*; do not reject if it includes that association". This interval rule is equivalent to the $\alpha$-level rule: False rejection corresponds to failure of the interval to contain the association predicted by *H* when *H* is correct; false acceptance corresponds to the interval containing that predicted association when *H* is incorrect. The $1-\alpha$ interval shows all association values for the association that have $p > \alpha$ and would *not* be rejected using the rule; thus, examining the entire interval can prevent naïve acceptance of *H* simply because it has $p > \alpha$.

There is now a vast literature documenting and cataloguing the many mistaken inferences in research reports caused by dichotomous testing and its terminology (e.g., see Amrhein et al. 2017, 2019a, 2019b; Gelman 2016, 2018; Gigerenzer 2004; Gigerenzer et al. 2004; Gigerenzer and Marewski 2015; Greenland et al. 2016b; Greenland 2017, 2023b; McShane and Gal 2017;

McShane et al. 2019, 2024; Ting and Greenland 2024). Foremost, the $\alpha$-level rule is usually called a "significance test", with $p \leq \alpha$ being called "statistically significant" and $p > \alpha$ "not significant". As discussed below, these tests and terms have been so extensively misused and misinterpreted that there have been calls for their complete abandonment, with replacement by interval estimates and by continuous *P*-values without reference to cut-offs (e.g., Amrhein et al. 2019a, 2019b; Campitelli 2019; McShane et al. 2019, 2024; Wasserstein et al. 2019; Rafi and Greenland 2020; Amrhein and Greenland 2022; Greenland et al. 2022; Greenland 2023b).

Misinterpretations aside, a more fundamental criticism of statistical tests is that there is usually no sound contextual rationale for basing a decision on a single statistical dichotomy. In this view, a study should seek to accurately report how it was conducted and what it observed in terms of estimates and *P*-values (Poole 1987; Rosnow and Rosenthal 1989; Cox and Donnelly 2011, Sec. 8.4.3; Amrhein et al. 2019a, 2019b; McShane et al. 2019, 2024; Rafi and Greenland 2020; Amrhein and Greenland 2022; Greenland et al. 2022). If a decision is needed, it should be based on the full information from the study and background, including the precise estimates and *P*-values available from each study and from meta-analyses, because judging the likely magnitude of error will be crucial in anticipating practical consequences.

*Problems in the terminology and conventions of significance testing*

Much misuse and misinterpretation of statistical testing can be traced to a synergy of the misuse of ordinary words for technical concepts that correspond poorly to their common meaning, and how the resulting misinterpretations create an illusion of simplicity or decisiveness of the results.

Notably, the original meaning of "accept/reject" terminology envisioned a situation in which the phrases "accept *H*" and "reject *H*" are only decisions to use *H* or not until better working hypotheses or more data become available. Unfortunately, when testing null hypotheses ($H_0$) of "no association" or "no effect", the phrases came to be misinterpreted as declarations about $H_0$ being correct or incorrect, as if the rule was an authoritative oracle saying the data favored $H_0$ if $p > 0.05$. This fallacious *nullistic* interpretation is encouraged by the term "acceptance" and has led to an enormous number of reports erroneously concluding there is no association or no effect simply because $p > \alpha$, when in fact their own estimates indicated nothing of the sort; see Greenland (2011, 2012), Rafi and Greenland (2020), and Greenland et al. (2022) for examples. In Table 1, $OR_{\text{pop}} = 1$ ($H_0$) has $p = 0.041$ and so would be "accepted" if $\alpha$ were 0.04 or less; yet $OR_{\text{pop}} = 6$ has an exact $p = 0.070$ and thus is more compatible with the data.

Equally misleading is the use of "significant" for $p \leq \alpha$ and "not significant" for $p > \alpha$, because this statistical decision rule has little or no relation to the practical significance of the results: Practical significance depends directly on the effect size and the *degree* of error. The confusion of statistical and practical significance has persisted even with the more precise labels of "statistically significant" and "not statistically significant", as in the originating literature. Remarkably, the confusion was criticized as far back as Boring (1919), and the mistake of misinterpreting "not statistically significant" as indicating no association was criticized even earlier by Pearson (1906). Adding to the confusion, a large segment of the 20$^{\text{th}}$-century literature (especially British) used "significance level" as a synonym for the *P*-value, but another large segment (especially American) used "significance level" as a synonym for the decision cut-off $\alpha$.

As mentioned earlier, another problem is that most of literature uses the term “null hypothesis” to refer to any statistical hypothesis *H*, despite that fact that in ordinary English the term “null” means zero or nothing. This misuse of “null” for any hypothesis has led to examining only hypotheses of “no association” or “no effect”, while neglecting reasonable alternatives to that null. The misleading terminology is compounded by calling any use of *P*-values or $\alpha$-level rules “null-hypothesis significance testing” (NHST).

Terminology aside, dichotomous testing reduces the *P*-value or interval to an intermediate computation in a statistical decision rule, which can be very misleading because it ignores and even obscures important data information. For example, *H* may be far from the associations most compatible with the data, even if $p > 0.05$. Again, in Table 1, the hypothesis $OR_{pop} = 1$ has $p = 0.041$, so would be “accepted” when using $\alpha = 0.04$; nonetheless, $OR_{pop} = 2$ has $p = 0.64$ and so is much more compatible with the tabulated data.

Yet another problem is that, although the $\alpha = 0.05$ cut-point was only casually mentioned by R.A. Fisher as a preference in his own research, it became a rigid convention enforced by some journals, despite many objections to 0.05 as a universal cut-off for “statistical significance” including by Fisher himself (Fisher 1955) and many explanations of why $\alpha$ should be determined on a study-specific basis (Neyman 1977; Mayo 2018; Lakens et al. 2018). Despite those explanations, some statisticians have campaigned for using $\alpha = 0.005$ as a new convention to reduce the frequency with which a correct *H* is rejected (“false-positive errors”) (Benjamin et al., 2018). This campaign ignores that some researchers are more concerned about the opposite error of accepting an incorrect *H* (“false-negative errors”), an error whose frequency is increased by using a smaller $\alpha$. Thus, lowering $\alpha$ to reduce incorrect rejections (false positives or Type-I errors) will increase incorrect acceptances (false negatives or Type-II errors) and increase the publication bias and the inflation of "significant" estimates discussed below (van Calster et al. 2018). Given the conflicting concerns, $\alpha = 0.05$ remains the convention in most venues.

*Publication bias and the illusion of "replication failure"*

Among the severe literature distortions produced by misuse and misinterpretations of statistics is *publication bias*, which usually refers to the preferential publishing of results that are "statistically significant" in the mistaken belief that those results are more worthy of attention. This bias results in published estimates that are on average inflated in size (farther away from the null than the actual underlying effects), as well as other distortions as illustrated for example in Fig. 1 of van Zwet and Cator (2021); see also Ioannidis (2008), Gelman (2018), and Gigerenzer (2018).

Once a "significant" association is reported, however, an opposite deflationary publication bias may arise in which "significant" results are no longer considered to be as interesting as "non-significant" results when the latter are misreported as in conflict with the original "significant" result. This leads to reviews that report conflict or "replication failure" among studies because some are "statistically significant" and others are not, when in fact there is no difference among the studies greater than that expected by chance (Rothman 1986; Amrhein et al. 2019b; Ting and Greenland 2023). To avoid such mistaken reporting of conflict, the studies must be compared directly using $P$-values and estimates for their differences (Greenland et al. 2016b, item 16).

*Error rates and power of binary statistical decisions*

The usual justification offered for statistical decision rules is that their use supposedly comes with guarantees about the rates of erroneous decisions (Neyman 1977; Mayo 2018). This rationale is fallacious when the assumptions used to derive the rules are uncertain, because actual error rates are extremely sensitive to plausible violations of random-sampling and randomization assumptions. Yet “significance tests” are routinely applied to studies for which no such assumption can be justified, and many bad decisions have ensued from this sort of misapplication.

Now consider the ideal situation in which all the needed assumptions have been fully enforced by the study design and execution. Then, when testing a correct hypothesis $H$, the “$p \leq \alpha$” rule will incorrectly reject $H$ (commit Type-I error) in no more than $100\alpha$% of possible random samples and “accept” $H$ in no less than $(1-\alpha)100$% of the other samples. Thus, $\alpha$ is often called the “false-rejection rate”, although this term is slightly inaccurate because $\alpha$ is not that rate but is instead a pre-specified cap (maximum) for the rate. Even under ideal conditions, the actual rate at which the test rejects a correct $H$ can be substantially less than $\alpha$ for exact tests in tables with small expected counts. And again, $\alpha$ does *not* refer to the probability that the actual decision taken is incorrect, nor does it account for uncertainties about auxiliary assumptions.

The rate at which the rule rejects an incorrect $H$ is called the *power* or “true-positive rate” of the test of $H$ and is inversely related to both $\alpha$ and how close $H$ is to being correct. To illustrate via extreme cases, if $\alpha$ were 0, we would never reject (i.e., the power would be zero), whereas if $\alpha$ were 1, we would always reject (i.e., the power would be 100%). Similarly, for a typical study using $\alpha = 0.05$, if $OR_{pop}$ were 1.01, the rejection rate for $H_0$: $OR_{pop} = 1$ would be little different from $\alpha$; but if $OR_{pop}$ were 100, the rejection rate for $H_0$ would approach 100%.

The rate at which the test fails to reject an incorrect $H$ is 1−power, often called the "false-negative" or Type-II error rate and denoted by $\beta$. It varies directly with $\alpha$ and how close $H$ is to being correct, hence to be precise we should denote it by $\beta(\alpha, OR_{pop})$ to show how it depends on the chosen $\alpha$ and the unknown correct odds ratio $OR_{pop}$.

*Problems with power*

When $\alpha = 0.05$, it is a common convention to design a study to have 80% power for some pre-specified alternative $H_1$ to $H_0$, so that $\beta = 0.20$; this convention makes the false-negative rate $\beta$ at least four times the false-positive rate. This disparity in error rates is rarely given any sound justification and is harmful if failure to reject $H_0$ when $H_1$ is correct is more costly than false rejection of $H_0$; in those cases, the roles of $H_0$ when $H_1$ should be reversed (Neyman 1977).

A deeper problem is that power is unknown because the correct association is unknown, which makes power computation in study design highly speculative. This problem may be partially addressed by graphing or tabulating power against different plausible values of the association, to produce a *power curve* showing the power of the $\alpha$-level test under different alternatives to the target hypothesis $H$.

Power is unnecessary and arguably irrelevant for data analysis because the statistical information it conveys is already visible in the more intuitively understandable form of point and interval estimates; hence the use of power should be limited to the design of further studies (Goodman and Berlin 1994; Hoenig and Heisey 2001; Senn 2002). Power can be highly misleading for interpreting the results; for example, data may show "non-significance" for $H_0$ with "high power" (e.g., $p > 0.05$ with power > 80%) and so be misinterpreted as supporting "no effect", when in fact the data may be much more compatible or conflict less with important effects (Greenland 2012; Greenland et al. 2016b).

*Multiple comparisons*

A highly controversial and subtle problem in the use of "confidence" and "significance" interpretations of intervals and *P*-values arises under the heading of *multiple comparisons* (also known as simultaneous inference). There are many versions of the problem that arise in practice, with many conflicting solutions and opinions about them. Greenland (2021b) provides an overview of the controversy that attempts to explain the sources of conflict and the theory behind the positions.

In one version of the problem, suppose we analyze a data set from a large random survey with the goal of deciding which one of 20 hypotheses $H_1, \ldots, H_{20}$ to recommend pursuing based on significance testing each one separately at the $\alpha = 0.05$ level. If all 20 hypotheses were correct and their *P*-values were statistically independent, the chance of getting one or more with $p \leq 0.05$ (and thus falsely rejected) would be $1-0.95^{20} = 64\%$, not the $1-0.95^{1} = 5\%$ we'd expect from testing only one correct hypothesis. This fact has led some statisticians to encourage use of what is called a *Bonferroni adjustment* in this problem; when applied to $\alpha$, it says we should employ $\alpha/20 = 0.0025$ or 0.25% as our cut-off. The chance of getting one or more hypotheses with $p \leq 0.0025$, and thus falsely rejected, would now be $1-0.9975^{20} = 4.9\%$.

Although this result may initially seem to justify a Bonferroni adjustment, it is very misleading for practical purposes because it takes no account of power, data dependencies, or error costs. Most importantly, the power to reject false hypotheses is drastically reduced by uniformly lowering the decision cut-off to $\alpha/20$. The power loss is further worsened because the tests use the same variables and data to decide each hypothesis, and so the tests cannot be independent. The extent of dependencies can be quite large, resulting in the chance of a false rejection among 20 correct hypotheses being far less than indicated by the independence-based formula of $1-(1-\alpha)^{20}$. To see this in an extreme example, suppose the tests were all perfectly positively correlated (so that they all reject or fail to reject together) and all the hypotheses were correct; the rate at which the Bonferroni procedure would make a false rejection would equal the new cut-off of 0.25%, not 5%. Among the consequences of having the actual false-rejection rate far below the cut-off is that the power to detect false hypotheses would be reduced to a few percent.

Considering costs, the researchers may be willing to accept the 64% chance of getting at least one false rejection if all 20 hypotheses are correct, in part because they may expect that some of the hypotheses are false and don't want to harm their power to find them. The analysis task should then be reframed as detecting which of the hypotheses are false among the total while minimizing both false detections and detection failures. There are many multiple-comparison methods far more sophisticated than Bonferroni that address such questions (Parmigiani and Inoue 2009; Efron 2010).

A deeper problem is determining whether a multiple-comparison method is justified at all. Many authors have argued that, if the analyst's goal is to separately describe the compatibility of different hypotheses with the data or to summarize data information about associations or effects (rather than make decisions), multiple-comparison procedures have no sound rationale (Rosenthal and Rubin 1985; Rothman 1990; Rubin 2024). Other authors argue that most settings in which multiplicity arises are best analyzed using Bayesian or empirical-Bayes methods (Efron 2010; Greenland 2021b).

**Conclusions**

This chapter has reviewed the basic elements of conventional statistics, focusing on how they assume severe oversimplifications that fail to align with subject-matter and research realities in studies of human populations. These failings lead to the overinterpretations evoked by traditional statistics in which a P-value indicates the "significance" of an observed association and an interval estimate provides a range of "confidence" for the actual association size. Safe use of statistical methods instead requires emphasizing that they provide inferences only in ideal cases, rather than "correct answers" for our actual imperfect applications. Problems can be reduced by rejecting statistical methods as oracles for scientific inferences, and instead using them to describe the relation of data to models for the behavior of the processes that generated the data. This reorientation includes

1) re-interpreting *P*-values as measures of compatibility of the data with hypothetical models for associations,

2) re-interpreting interval estimates as ranges for association sizes that have high compatibility with the data under explicit assumptions about the processes that generated the data, and
3) describing and accounting for the uncertainties in the assumptions used to compute the statistics before using the statistics to inform inferences or decisions.

**Cross-references**

Bayesian Methods in Epidemiology

Causal Directed Acyclic Graphs

Causal Reasoning and Inference in Epidemiology

Confounding and Interaction

Regression Methods for Epidemiological Analysis

Sensitivity Analysis and Bias Analysis

**References**

Agresti, A.A. (2018). *Introduction to Categorical Data Analysis*, 3rd edn. Wiley, New York.

Amrhein, V., Korner-Nievergelt, F., Roth,T. (2017). The earth is flat (p> 0.05): significance thresholds and the crisis of unreplicable research. *Peer J*, 5, e3544, https://peerj.com/articles/3544/

Amrhein, V., Greenland, S., McShane, B. (2019a). Retire statistical significance. *Nature*, 567, 305-307, https://media.nature.com/original/magazine-assets/d41586-019-00857-9/d41586-019-00857-9.pdf

Amrhein, V., Trafimow, D., and Greenland, S. (2019b). Inferential statistics as descriptive statistics: There is no replication crisis if we don't expect replication. *The American Statistician*, 73 supplement 1, 262-270, open access at www.tandfonline.com/doi/pdf/10.1080/00031305.2018.1543137

Amrhein, V., Greenland, S. (2022). Discuss practical importance of results based on interval estimates and p-value functions, not only on point estimates and null p-values. *Journal of Information Technology*, 37(3), 316-320, doi:10.1177/02683962221105904

Bedrick, E. J., Christensen, R., Johnson, W. (1996). A new perspective on generalized linear models. *Journal of the American Statistical Association*, 91, 1450-1460.

Benjamin, D.J., Berger, J.O., Johannesson, M., Nosek, B.A., Wagenmakers, E., Berk, R., et al. (2018). Redefine statistical significance. *Nature Human Behavior*, 2, 6-10.

Boring, E.G. (1919). Mathematical vs. scientific significance. *Psychological Bulletin*, 16, 335-338.

Box, G.E.P. (1980). Sampling and Bayes inference in scientific modeling and robustness (with discussion). *Journal of the Royal Statistical Society*, Series A, 143, 383-430.

Box, G.E.P. (1990). The unity and diversity of probability: Comment. *Statistical Science*, 5(4), 448-449.

Campitelli, G. (2019). Retiring statistical significance from psychology and expertise research. *Journal of Expertise Research*, 2(4), 217-223.

Carroll, R.J., Ruppert, D., Stefanski, L.A., & Crainiceanu, C. (2006) *Measurement Error in Nonlinear Models*, 2nd edn. Chapman & Hall, Boca Raton, FL.

Clark, D.R. (2025). Credibility as data augmentation. *CAS E-Forum*, 03 April, https://eforum.casact.org/article/132224-credibility-as-data-augmentation

Cole S.R., Edwards, J.K., Greenland, S. (2021). Surprise! *American Journal of Epidemiology*, 190(2):191-193. https://doi.org/10.1093/aje/kwaa136

Cox, D. R., Wermuth, N. (1992). A comment on the coefficient of determination for binary responses. *The American Statistician*, 46, 104.

Cox, D.R., Donnelley, C. (2011). *Principles of Applied Statistics*. Cambridge University Press, New York, sec. 8.4.3.

Dahabreh, I.J., Robertson, S.E., Steingrimsson, J.A. (2024). Learning about treatment effects in a new target population under transportability assumptions for relative effect measures. *European Journal of Epidemiology*, 39, 957-965, https://doi.org/10.1007/s10654-023-01067-4

Deaton, A., Cartwright, N. (2018). Understanding and misunderstanding randomized controlled trials (with discussion). *Social Science & Medicine*, 210, 2–90.

DeFinetti, B. (1975). *The Theory of Probability*, vol. 2. New York: Wiley.

Discacciati, A., Orsini, N., Greenland, S. (2015). Approximate Bayesian logistic regression via penalized likelihood by data augmentation. *The Stata Journal*, 15(3), 712-736. https://www.stata-journal.com/article.html?article=st0400

Efron, B. (2010). *Large-Scale Inference: Empirical Bayes Methods for Estimation, Testing, and Prediction*. New York: Cambridge University Press.

Fisher, R.A. (1934). *Statistical Methods for Research Workers*, 5th edn. London, Oliver Boyd.

Fisher, R.A. (1936). Has Mendel's work been rediscovered? *Annals of Science*, 1, 115-137.

Fisher, R. A. (1955). Statistical methods and scientific induction. *Journal of the Royal Statistical Society*, Series B, 17(1), 69-78.

Fox, M.P., MacLehose, R.F., Lash, T.L. (2021) *Applying Quantitative Bias Analysis to Epidemiologic Data*, 2nd edn. New York, Springer

Ganesh, S., D'Souza, D.C. (2022). Cannabis and psychosis: recent epidemiological findings continuing the "causality debate." *American Journal of Psychiatry*, 179, 8-10.

Gelman, A. (2016). The problems with *P*-values are not just with *P*-values. *The American Statistician*, 70, online discussion at https://sites.stat.columbia.edu/gelman/research/published/asa_pvalues.pdf

Gelman, A. (2018). The failure of null hypothesis significance testing when studying incremental changes, and what to do about it. *Personality and Social Psychology Bulletin*, 44(1), 16-23

Gelman, A., Loken, (2014a). The statistical crisis in science. *American Scientist*, 102, 460–465.

Gelman, A., Loken, (2014b). The AAA tranche of subprime science. *Chance*, 27(1), 51-56.

Gelman, A., Carlin, J. B., Stern, H. S., Dunson, D. B., Vehtari, A., Rubin, D. B. (2013). *Bayesian Data Analysis*, third edition. London: Chapman and Hall.

Gigerenzer, G. (2004). Mindless statistics. *Journal of Socioeconomics*, 33, 567–606.

Gigerenzer, G., Marewski, J.N. (2015). Surrogate Science: The idol of a universal method for scientific inference. *Journal of Management*, 41, 421-440.

Gigerenzer, G. (2018). Statistical rituals: the replication delusion and how we got there. *Advances in Methods and Practices in Psychological Science*. 1(2), 198-218

Gigerenzer, G., Krauss, S., Vitouch, O (2004). The null ritual: what you always wanted to know about significance testing but were afraid to ask. In: Kaplan, D., ed. *The Sage handbook of quantitative methodology for the social sciences.* Thousand Oaks: Sage Publications, 391-408.

Glymour, M.M., Greenland, S. (2008). Causal diagrams. Ch. 12 in Rothman, K. J., Greenland, S., Lash, T.L. (2008), eds. *Modern Epidemiology*, 3rd edn. Philadelphia: Lippincott-Wolters-Kluwer.

Good, I.J. (1983). *Good Thinking*. Minneapolis: University of Minnesota Press, Ch. 3, 20-21.

Goodman, S.N., Berlin, J. (1994). The use of predicted confidence intervals when planning experiments and the misuse of power when interpreting results. *Annals of Internal Medicine*, 121, 200-206.

Greenland, S. (2005). Multiple-bias modeling for analysis of observational data (with discussion). *Journal of the Royal Statistical Society*, Series A, 168, 267-308.

Greenland, S. (2008a). Analysis of polytomous exposures and outcomes. Ch. 17 in Rothman, K. J., Greenland, S., Lash, T.L. (2008), eds. *Modern Epidemiology*, 3rd edn. Philadelphia: Lippincott-Wolters-Kluwer, 303-327.

Greenland, S. (2008b). Introduction to Bayesian statistics. Ch. 18 in Rothman, K. J., Greenland, S., Lash, T.L. (2008), eds. *Modern Epidemiology*, 3rd edn. Philadelphia: Lippincott-Wolters-Kluwer, 328-344.

Greenland, S. (2009). Accounting for uncertainty about investigator bias: disclosure is informative. *American Journal of Epidemiology and Community Health*, 63, 593–598. doi:10.1136/jech.2008.084913

Greenland, S. (2011). Null misinterpretation in statistical testing and its impact on health risk assessment. Preventive Medicine, 53, 225-228.

Greenland, S. (2012). Nonsignificance plus high power does not imply support for the null over the alternative. *Annals of Epidemiology*, 22, 364–368.

Greenland, S. (2017). The need for cognitive science in methodology. *American Journal of Epidemiology*, 186, 639-645, doi:10.1093/aje/kwx259.

Greenland, S. (2021a). Noncollapsibility, confounding, and sparse-data bias. Part 2: What should researchers make of persistent controversies about the odds ratio? *Journal of Clinical Epidemiology*, 139, 264-268.

Greenland, S. (2021b). Analysis goals, error-cost sensitivity, and analysis hacking: essential considerations in hypothesis testing and multiple comparisons. *Pediatric and Perinatal Epidemiology*, 35, 8-23, doi:10.1111/ppe.12711

Greenland, S. (2022). The causal foundations of applied probability and statistics. Ch. 31 in: Dechter, R., Halpern, J., and Geffner, H., eds. *Probabilistic and Causal Inference: The Works of Judea Pearl*. ACM Books, no. 36, 605-624, https://dl.acm.org/doi/10.1145/3501714.3501747, corrected version at https://arxiv.org/abs/2011.02677

Greenland, S. (2023a). Divergence vs. decision P-values: A distinction worth making in theory and keeping in practice (with discussion). *Scandinavian Journal of Statistics*, 50(1), 1-35, doi:10.1111/sjos.12625, discussion 50(3), 899-933, corrigendum 51(1), 425

Greenland, S. (2023b). Connecting simple and precise p-values to complex and ambiguous realities (includes rejoinder to comments on "Divergence vs. decision P-values"). *Scandinavian*

*Journal of Statistics*, 50(3), 899-914, https://arxiv.org/abs/2304.01392, https://onlinelibrary.wiley.com/doi/10.1111/sjos.12645

Greenland, S. (2024a). Regression methods for epidemiological analysis. Chapter 17 in: Ahrens, W., and Pigeot, I. (eds.). *Handbook of Epidemiology*, 3rd ed. New York: Springer. https://link.springer.com/referenceworkentry/10.1007/978-1-4614-6625-3_17-1

Greenland, S. (2024b). Sensitivity analysis and bias analysis. Chapter 60 in: Ahrens, W., and Pigeot, I. (eds.). *Handbook of Epidemiology*, 3rd edn. New York: Springer, https://link.springer.com/referenceworkentry/10.1007/978-1-4614-6625-3_60-1

Greenland, S. (2025). Some ways to make regression modeling more helpful than misleading. *Statistics in Medicine*, 44, in press, https://doi.org/10.1002/sim.10313

Greenland, S., Mansournia, M.A. (2015). Limitations of individual causal models, causal graphs, and ignorability assumptions, as illustrated by random confounding and design unfaithfulness. *European Journal of Epidemiology*, 30, 1101-1110, https://link.springer.com/article/10.1007%2Fs10654-015-9995-7

Greenland, S., Robins, J.M. (2009). Identifiability, exchangeability and confounding revisited. *Epidemiologic Perspectives & Innovations*, 6:4, http://www.epi-perspectives.com/content/6/1/4

Greenland, S., Schlesselman, J. J., Criqui, M. H. (1986). the fallacy of employing standardized regression coefficients and correlations as measures of effect. *American Journal of Epidemiology*, 123, 203-208.

Greenland, S., Maclure, M., Schlesselman, J. J., Poole, C., Morgenstern, H. (1991). Standardized regression coefficients: a further critique and a review of alternatives. *Epidemiology*, 2, 387-392.

Greenland, S., Pearl, J., Robins, J. M. (1999). Causal diagrams for epidemiologic research. *Epidemiology*, 10, 37-48.

Greenland, S., Mansournia, M.A., Altman, D.G. (2016a). Sparse-data bias: A problem hiding in plain sight. *British Medical Journal*, 353:i1981, 1-6, https://www.bmj.com/content/352/bmj.i1981.

Greenland, S., Mansournia, M., Joffe, M. (2022). To curb research misreporting, replace significance and confidence by compatibility. *Preventive Medicine*, 164, https://www.sciencedirect.com/science/article/pii/S0091743522001761

Greenland, S., Senn, S.J., Rothman, K.J., Carlin, J.C., Poole, C., Goodman, S.N., Altman, D.G. (2016b). Statistical tests, confidence intervals, and power: A guide to misinterpretations. *The American Statistician*, 70(suppl.1), 1-12, open access at https://amstat.tandfonline.com/doi/suppl/10.1080/00031305.2016.1154108/suppl_file/utas_a_1154108_sm5368.pdf

Gustafson, P. (2015). *Bayesian inference for partially identified models: Exploring the limits of limited data*. Boca Raton, FL. CRC Press.

Gutierrez, S., Glymour, M.M., Davey Smith, G. (2025). Evidence triangulation in health research. European Journal of Epidemiology, 40, 743-757.

Haller, H., Krauss, S. (2002). Misinterpretations of significance: A problem students share with their teachers? *Methods of Psychological Research*, 7(1), 1-20.

Harrell, F.E. (2015). *Regression Modeling Strategies*. Springer, New York.

Hernán, M.A., Cole, S.R. (2009). Causal diagrams and measurement bias. *American Journal of Epidemiology*, 170, 959-962.

Hernán, M.A., Robins, J.M. (2025). *Causal Inference: What If?* Chapman Hall, New York

Hill, A.B. (1965). The environment and disease: association or causation? *Proceedings of the Royal Society of Medicine*, 58, 295-300.

Hoenig, J.M., Heisey, D.M. (2001). The abuse of power: the pervasive fallacy of power calculations for data analysis. *The American Statistician*, 55, 19-24.

Hosmer, D.W., Lemeshow, S., Sturtevant, R.X. (2013). *Applied Logistic Regression*, 3rd edn. New York: Wiley.

Ioannidis, J.P.A. (2008). Why most discovered true associations are inflated. *Epidemiology*, 19, 640-648. DOI: 10.1097/EDE.0b013e31818131e7.

Jewell, N.P. (2004). *Statistics for Epidemiology*. Chapman & Hall/CRC, New York.

Kummerfeld, E., Jones, G.L. (2023). One data set, many analysts: Implications for practicing scientists. Frontiers in Psychology, 14, doi:10.3389/fpsyg.2023.1094150

Lakens, D., Adolfi, F. G., Albers, C., Anvari, F., Apps, M. A., Argamon, S. E., Baguley, T., Becker, R. B., Benning, S. D., Bradford, D. E., Buchanan, E. M., et al. (2018). Justify your alpha. *Nature Human Behaviour*, 2, 168-171

Lash, T.L., VanderWeele, T.J., Haneuse, S., Rothman, K.J. (2021). *Modern Epidemiology*, 4th edn. Philadelphia: Lippincott-Wolters-Kluwer.

Lecoutre, M.-P., Poitevineau, J., Lecoutre, B. (2003). Even statisticians are not immune to misinterpretations of Null Hypothesis Tests. *International Journal of Psychology*, 38, 37-45

Little R.J.A., Rubin D.B. (2019). *Statistical Analysis with Missing Data*, 3rd edn. New York: Wiley.

Mayo, D. (2018). *Statistical Inference as Severe Testing*. Cambridge University Press, New York.

McElreath, R. (2020). *Statistical Rethinking: A Bayesian Course with Examples in R and STAN*, 2nd edn. Chapman Hall/CRC, New York.

McShane, B., Gal, D. (2017). Statistical significance and the dichotomization of evidence. *Journal of the American Statistical Association*, 112,

McShane, B.B., Gal, D., Gelman, A., Robert, C., Tackett, J.L. (2019). Abandon statistical significance. *The American Statistician*, 73(suppl.1), 99-105, doi:10.1080/00031305.2018.1505655

McShane, B.B., Bradlow, E.T., Lynch Jr., J.G., Meyer, R.J. (2024). "Statistical significance" and statistical reporting: moving beyond binary. *Journal of Marketing*, 88(3), 1-19

Neyman, J. (1977). Frequentist probability and frequentist statistics. *Synthese*, 36, 97-131.

Parmigiani, G., Inoue, L. (2009). *Decision Theory: Principles and Approaches*. New York: Wiley.

Pearl, J. (2009). *Causality: Models, Reasoning and Inference*. 2nd edn. Cambridge: Cambridge University Press.

Pearl, J., Glymour, M., Jewell, N.P. (2016). *Causal Inference in Statistics: A Primer*. New York: Wiley.

Pearson, K. (1900). On the criterion that a given system of deviations from the probable in the case of a correlated system of variables is such that it can be reasonably supposed to have arisen from random sampling. *London, Edinburgh, and Dublin Philosophical Magazine and Journal of Science*, 50, 157-175.

Pearson, K. (1906). Note on the significant or non-significant character of a subsample drawn from a sample. *Biometrika*, 5, 181-183.

Poole, C. (1987). Beyond the confidence interval. *Journal of the American Public Health Asssociation*, 77, 195-199.

Rafi, Z., Greenland, S. (2020). Semantic and cognitive tools to aid statistical science: Replace confidence and significance by compatibility and surprise. *BMC Medical Research Methodology*, 20, 244. doi:10.1186/s12874-020-01105-9

Rosenthal, R., Rubin, D.B. (1979). A note on percent variance explained as a measure of importance of effects. *Journal of Applied Social Psychology*, 9, 395-396.

Rosenthal, R., Rubin, D.B. (1985). Statistical analysis: Summarizing evidence versus establishing facts. *Psychological Bulletin*, 97(3), 527-529

Rosnow, R.L., Rosenthal, R. (1989). Statistical procedures and the justification of knowledge in psychological science. *American Psychologist*, 44, 1276-1284.

Rothman, K.J. (1986). Significance questing. *Annals of Internal Medicine*, 105, 445-447

Rothman, K.J. (1990). No adjustments are needed for multiple comparisons. *Epidemiology*, 1, 43-46. https://www.jstor.org/stable/20065622.

Rovetta, A., Piretta, L., Mansournia, M.A. (2025). P-values and confidence intervals as compatibility measures: Guidelines for interpreting statistical studies in clinical research. *The Lancet Regional Health Southeast Asia*, 33, 1-3, https://www.thelancet.com/journals/lansea/article/PIIS2772-3682(25)00005-8/fulltext

Rubenstein, S. (2009). A new low in drug research: 21 fabricated studies. *The Wall Street Journal*, 11 March. https://www.wsj.com/articles/BL-HEB-8367, 33, 100534

Rubin, M. (2024). Inconsistent multiple testing corrections: The fallacy of using family-based error rates to make inferences about individual hypotheses. *Methods in Psychology*, 10, 100140, doi:10.1016/j.metip.2024.100140

Senn, S. (2002). Power is indeed irrelevant in interpreting completed studies. *BMJ* 325, 1304.

Shafer, G. (2020). On the nineteenth-century origins of significance testing and p-hacking. *Game-Theoretic Probability and Finance Project*, Working Paper 55, rev. 11 June 2020, http://probabilityandfinance.com/articles/55.pdf

Silberzahn, R., Uhlmann, E. L., Martin, D. P., Anselmi, P., Aust, F., Awtrey, E., Nosek, B. A. (2018). Many analysts, one data set: Making transparent how variations in analytic choices affect results. *Advances in Methods and Practices in Psychological Science*, 1, 337–356. doi:10.1177/2515245917747646

Skellam, J.G. (1969). Models, inference, and strategy. *Biometrics*, 25(3), 457-475.

Stigler, S.M. (1986). *The History of Statistics: The Measurement of Uncertainty before 1900*. Harvard University Press, Cambridge, MA.

Sullivan, S., Greenland, S. (2013). Bayesian regression in SAS software. *International Journal of Epidemiology*, 42, 308-317, https://academic.oup.com/ije/article/42/1/308/698250. Erratum (2014), 43, 1667–1668, https://academic.oup.com/ije/article/43/5/1667/2949610

Ting, C., Greenland, S. (2024). Forcing a deterministic frame on probabilistic phenomena: A communication blind spot in media coverage of the "replication crisis." *Science Communication*, 46, 672-684, https://journals.sagepub.com/doi/10.1177/10755470241239947.

Tukey, J. W. (1954). Causation, regression, and path analysis. In: Kempthorne, O., ed. *Statistics and Mathematics in Biology*. Ames: Iowa State College Press, 35-66.

van Calster, B., Steyerberg, E.W., Collins, G.S., Smits, T. (2018). Consequences of relying on statistical significance: some illustrations. *European Journal of Clinical Investigation*, 48(5), e12912, https://doi.org/10.1111/eci.12912

van Zwet, E.W., Cator, E.A. (2021). The significance filter, the winner's curse and the need to shrink. *Statistica Neerlandica*, 75, 437-452. DOI: 10.1111/stan.12241.

VanderWeele, T.J. (2015). *Explanation in Causal Inference: Methods for Mediation and Interaction*. New York: Oxford University Press.

Wasserstein, R.L., Lazar, N.A. (2016). The ASA's statement on p-values: Context, process, and purpose. *The American Statistician*, 70, 129-133, https://doi.org/10.1080/00031305.2016.1154108

Wasserstein, R.L., Schirm, A.L., Lazar, N.A. (2019). Moving to a world beyond “p<0.05”. *The American Statistician*, 73, 1-19, doi:10.1080/00031305.2019.1583913